\documentstyle[sprocl]{article}

\input{psfig}

\def\be{\begin{equation}}
\def\ee{\end{equation}}
\def\bea{\begin{eqnarray}}
\def\eea{\end{eqnarray}}

\begin{document}

\title{ALGEBRAIC MODEL OF BARYON STRUCTURE}

\author{R. BIJKER}

\address{ICN-UNAM, A.P. 70-543, 04510 M\'exico, D.F., M\'exico} 

\author{A. LEVIATAN}

\address{Racah Institute of Physics, The Hebrew University,
Jerusalem 91904, Israel}

\maketitle

\abstracts{We discuss properties of baryon resonances 
belonging to the $N$, $\Delta$, $\Sigma$, $\Lambda$, $\Xi$ 
and $\Omega$ families in a collective string-like model 
for the nucleon, in which the radial excitations are interpreted 
as rotations and vibrations of the string configuration. 
We find good overall agreement with the available data. 
The main discrepancies are found for low lying $S$-wave states, 
in particular $N(1535)$, $N(1650)$, $\Sigma(1750)$, 
$\Lambda^*(1405)$, $\Lambda(1670)$ and $\Lambda(1800)$.} 

\section{Introduction}

The development of dedicated experimental facilities to probe 
the structure of hadrons in the nonperturbative region of QCD 
with far greater precision than before has stimulated us to 
reexamine hadron spectroscopy in a novel approach in which both 
internal (spin-flavor-color) and space degrees of freedom of 
hadrons are treated algebraically. The new ingredient is the 
introduction of a space symmetry or spectrum generating algebra 
for the radial excitations which for baryons was taken as 
$U(7)$ \cite{BIL1}. The algebraic approach unifies the harmonic 
oscillator quark model with collective string-like models of baryons. 

In this contribution we present an analysis of the mass 
spectrum and strong couplings of both nonstrange and strange baryon 
resonances in the framework of a collective string-like $qqq$ model 
in which the radial excitations are treated as rotations and 
vibrations of the strings. The algebraic structure of the model 
enables us to obtain transparent results (mass formula, selection 
rules and decay widths) that can be used to analyze and interpret 
the experimental data, and look for evidence of the existence of 
unconventional ({\it i.e.} non $qqq$) configurations of quarks and 
gluons, such as hybrid quark-gluon states $qqq$-$g$ or multiquark 
meson-baryon bound states $qqq$-$q\overline{q}$. 

\section{Mass spectrum}

We consider baryons to be built of three constituent parts which 
are characterized by both internal and radial (or spatial) degrees 
of freedom. The internal degrees of freedom are described by the 
usual spin-flavor (${\rm sf}$) and color (${\rm c}$) algebras 
$SU_{\rm sf}(6) \otimes SU_{\rm c}(3)$. The radial degrees of 
freedom for the relative motion of the three constituent parts are 
taken as the Jacobi coordinates, which are treated algebraically 
in terms of the spectrum generating algebra of $U(7)$ \cite{BIL1}. 
The full algebraic structure is obtained by combining the radial 
part with the internal spin-flavor-color part 
\bea
{\cal G} \;=\; U(7) \otimes SU_{\rm sf}(6) \otimes SU_{\rm c}(3) ~, 
\eea
in such a way that the total baryon wave function is antisymmetric. 

For the radial part we consider a collective (string-like) model 
of the nucleon in which the baryons are interpreted as rotational 
and vibrational excitations of an oblate symmetric top \cite{BIL1}. 
The spectrum consists of a series of vibrational excitations 
labeled by $(v_1,v_2)$, and a tower of rotational excitations built  
on top of each vibration. The occurrence of linear Regge 
trajectories suggests to add, in addition to the vibrational 
frequencies $\kappa_1$ and $\kappa_2$, a term linear in $L$. 
The slope of these trajectories is given by $\alpha$. 
For the spin-flavor part of the mass operator we use a 
G\"ursey-Radicati form \cite{GR}.
These considerations lead to a mass formula for 
nonstrange and strange baryons of the form \cite{BIL2}  
\bea
M^2 &=& M^2_0 + \kappa_1 \, v_1 + \kappa_2 \, v_2 + \alpha \, L 
+ a \, \Bigl[ \langle \hat C_2(SU_{ \rm sf}(6)) \rangle - 45 \Bigr]
\nonumber\\
&& + b \, \Bigl[ \langle \hat C_2(SU_{\rm f}(3)) \rangle -  9 \Bigr]
   + c \, \Bigl[ S(S+1) - \frac{3}{4} \Bigr] 
\nonumber\\
&& + d \, \Bigl[ Y - 1 \Bigr] + e \, \Bigl[ Y^2 - 1 \Bigr] 
   + f \, \Bigl[ I(I+1) - \frac{3}{4} \Bigr] ~.
\label{massformula}
\eea
The coefficient $M^2_0$ is determined by the nucleon mass 
$M_{0}^{2}=0.882$ GeV$^2$. The remaining nine coefficients are 
obtained in a simultaneous fit to 48 three and four star 
resonances which have been assigned as octet and decuplet states. 
We find a good overall description of both positive and negative 
baryon resonances of the $N$, $\Delta$, $\Sigma$, $\Lambda$, $\Xi$ 
and $\Omega$ families with an r.m.s. deviation of $\delta=33$ MeV 
\cite{BIL2} to be compared with $\delta=39$ MeV in a fit to 25 $N$ 
and $\Delta$ resonances \cite{BIL1}. 
There is no need for an additional energy shift for the positive 
parity states and another one for the negative parity states, 
as in the relativized quark model \cite{rqm}. 

The three resonances that were assigned as singlet states (and were 
not included in the fitting procedure) show a deviation of about 100 
MeV or more from the data: the $\Lambda^*(1405)$, $\Lambda^*(1520)$ 
and $\Lambda^*(2100)$ resonances are overpredicted by 236, 121 and 97 
MeV, respectively. An additional energy shift for the singlet states 
(without effecting the masses of the octet and decuplet states) can 
be obtained by adding to the mass formula of Eq.~(\ref{massformula}) 
a term $\Delta M^2$ that only acts on the singlet states. However, 
since $\Lambda^*(1405)$ and $\Lambda^*(1520)$ are spin-orbit partners, 
their mass splitting of 115 MeV cannot be reproduced by such a 
mechanism. In principle, this splitting can be obtained from a 
spin-orbit interaction, but the rest of the baryon spectra shows 
no evidence for such a large spin-orbit coupling. A more likely 
explanation is the proximity of the $\Lambda^*(1405)$ resonance to 
the $N \overline{K}$ threshold (see next section). 

A common feature to all $q^3$ quark models is the occurrence of 
missing resonances. In a recent three-channel analysis by the 
Zagreb group evidence was found for the existence of a third $P_{11}$ 
nucleon resonance at $1740 \pm 11$ MeV \cite{Zagreb}. The first 
two $P_{11}$ states at $1439 \pm 19$ MeV and $1729 \pm 16$ MeV 
correspond to the $N(1440)$ and $N(1710)$ resonances of the PDG 
\cite{PDG}. It is tempting to assign the extra resonance as one 
of the missing resonances \cite{CLRS}. In the present calculation 
it is associated with the $^{2}8_{1/2}[20,1^+]$ configuration and 
appears at 1713 MeV, compared to 1880 MeV in the relativized 
quark model (RQM) \cite{rqm} (see Table~\ref{missing}). 

\begin{table}
\centering
\caption[]{Masses of the first three $P_{11}$ states in MeV}
\vspace{10pt}
\label{missing}
\begin{tabular}{cccc}
\hline
& & & \\
PDG \protect\cite{PDG} & Zagreb \protect\cite{Zagreb} & 
RQM \protect\cite{CLRS} & present \protect\cite{BIL2} \\
& & & \\
\hline
& & & \\
$N(1440)$ & $1439 \pm 19$ & 1540 & 1444 \\
$N(1710)$ & $1729 \pm 16$ & 1770 & 1683 \\
          & $1740 \pm 11$ & 1880 & 1713 \\
& & & \\
\hline
\end{tabular}
\end{table}

A recent analysis of new data on kaon photoproduction \cite{Tran} 
has shown evidence for a $D_{13}$ resonance at 1895 MeV 
\cite{Mart}. In the present calculation, there are several 
possible assignments \cite{BIL2}. 
The lowest state that can be assigned to this new resonance 
is a vibrational excitation $(v_1,v_2)=(0,1)$ with 
$^{2}8_{3/2}[56,1^-]$ at 1847 MeV. This state belongs to 
the same vibrational band as the $N(1710)$ resonance. 
In the relativized quark model 
a $D_{13}$ state has been predicted at 1960 MeV \cite{rqm}. 

\section{Strong couplings}

Decay processes are far more sensitive to details in the baryon 
wave functions than are masses. Here we consider strong decays 
of baryons by the emission of a pseudoscalar meson
\bea
B \rightarrow B^{\prime} + M ~, 
\eea
in an elementary emission model \cite{BIL2}. The calculation 
of the strong decay widths involves a phase space factor, a 
spin-flavor matrix element and a spatial matrix element which is 
obtained in the collective model by folding with a distribution 
function over the volume of the nucleon. The calculations are 
carried out in the rest frame of the decaying resonance. The 
transition operator that induces the strong decay is determined 
in a fit to the $N \pi$ and $\Delta \pi$ channels which are 
relatively well known \cite{strong}. It is important to stress 
that in the present analysis the same transition operator is used 
for {\em all} resonances and {\em all} decay channels. 

The calculated decay widths are primarily due to spin-flavor 
symmetry and phase space. $N$ and $\Delta$ resonances decay 
predominantly into the $\pi$ channel, and strange resonances 
mainly into the $\pi$ and $\overline{K}$ channel. Phase space 
factors suppress the $\eta$ and $K$ decays. 
The use of the collective form factors introduces a power-law 
dependence on the meson momentum, compared to an exponential 
for harmonic oscillator form factors. 
In general, our results for the strong decay widths are in fair 
overall agreement with the available data, which shows that the 
combination of a collective string-like $qqq$ model of baryons 
and an elementary emission model for the decays can account for 
the main features of the data. 
As an example, in Table~\ref{del} we present the strong decays 
of three and four star $\Delta$ resonances. 

\begin{table}
\centering
\caption[]{Strong decay widths of three and four star delta 
resonances in MeV. For the $\eta$ mesons we introduce a mixing 
angle $\theta_P=-23^{\circ}$ between the octet 
and singlet mesons.  
The experimental values are taken from \protect\cite{PDG}. 
Decay channels labeled by -- are below threshold.}
\label{del} 
\vspace{10pt}
\begin{tabular}{lccccc}
\hline
& & & & & \\
Baryon & $N \pi$ & $\Sigma K$ 
& $\Delta \pi$ & $\Delta \eta$ & $\Sigma^*(1385) K$ \\
& & & & & \\
\hline
& & & & & \\
$\Delta(1232)P_{33}$ & 116 & -- & -- & -- & -- \\
& $119 \pm 5$ & & & & \\
$\Delta(1600)P_{33}$ & 108 & -- & 25 & -- & -- \\
& $61 \pm 32$ & & $193 \pm 76$ & & \\
$\Delta(1620)S_{31}$ &  16 & -- & 89 & -- & -- \\
& $38 \pm 11$ & & $68 \pm 26$ & & \\
$\Delta(1700)D_{33}$ &  27 & 0 & 144 & -- & -- \\
& $45 \pm 21$ & & $135 \pm 64$ & & \\
$\Delta(1905)F_{35}$ &   9 & 1 &  45 & 1 & 0 \\
& $36 \pm 20$ & & $< 45 \pm 45$ & & \\
$\Delta(1910)P_{31}$ &  42 &   2 &   4 & 0 & 0 \\
& $52 \pm 19$ & & & & \\
$\Delta(1920)P_{33}$ & 22 &   1 &  29 & 1 & 0 \\
& $28 \pm 19$ & & & & \\
$\Delta(1930)D_{35}$ &   0 &   0 &   0 &   0 &   0 \\
& $53 \pm 23$ & & & & \\
$\Delta(1950)F_{37}$ & 45 &   6 &  36 &   2 & 0 \\
& $120 \pm 14$ & & $80 \pm 18$ & & \\
$\Delta(2420)H_{3,11}$ & 12 &  4 &  11 &   2 &   1 \\
& $40 \pm 22$ & & & & \\
& & & & & \\
\hline
\end{tabular}
\end{table}

There are, however, a few exceptions which could indicate evidence 
for the importance of degrees of freedom which are outside the 
present $qqq$ model of baryons. The $\eta$ decays of octet baryons 
show an unusual behavior: the $S$-wave states $N(1535)$, 
$\Sigma(1750)$ and $\Lambda(1670)$ are found experimentally to have 
a large branching ratio to the $\eta$ channel with partial decay 
widths of $74 \pm 39$, $39 \pm 28$ and $9 \pm 5$ MeV, respectively 
\cite{PDG}. In our calculation these resonances are assigned as 
octet partners and only differ in their flavor content. The 
smallness of the calculated $\eta$ widths ($<0.5$ MeV) is mainly 
due to the available phase space.  
The results of this analysis suggest that the observed $\eta$ 
widths are not due to a conventional $qqq$ state, but may rather 
indicate evidence for the presence of a state in the same mass 
region of a more exotic nature, such as a quasi-molecular $S$-wave 
resonance $qqq$-$q\overline{q}$ just below or above threshold, 
bound by Van der Waals type forces \cite{Kaiser} (for example $N\eta$, 
$\Sigma \eta$ or $\Lambda \eta$). 

The decay of $^{4}8[70,L^P]$ $\Lambda$ states into the 
$N \overline{K}$ channel is forbidden by a spin-flavor selection 
rule \cite{BIL2} which is similar to the Moorhouse selection rule 
in electromagnetic couplings. Therefore, 
the calculated $N \overline{K}$ widths of $\Lambda(1800)$, 
$\Lambda(1830)$ and $\Lambda(2110)$ vanish, whereas all of 
them have been observed experimentally \cite{PDG}. 
The $\Lambda(1800)S_{01}$ state has large decay width into 
$N \overline{K}^{\,*}(892)$ \cite{PDG}. Since the mass of 
the resonance is just around the threshold of this channel, 
this could indicate a coupling with a quasi-molecular $S$ wave. 
The $N \overline{K}$ width of $\Lambda(1830)$ is 
relatively small ($6 \pm 3$ MeV), and hence in qualitative 
agreement with the selection rule. The situation for the 
the $\Lambda(2110)$ resonance is unclear. 

The $\Lambda^*(1405)$ resonance has a anomalously large decay width 
($50 \pm 2$ MeV) into $\Sigma \pi$. This feature emphasizes its 
quasi-molecular nature due to the proximity of the $N \overline{K}$ 
threshold. It has been shown \cite{Arima} that the inclusion of the 
coupling to the $N \overline{K}$ and $\Sigma \pi$ decay channels 
produces a downward shift of the $qqq$ state toward or even below 
the $N \overline{K}$ threshold. In a chiral meson-baryon Lagrangian 
approach with an effective coupled-channel potential the 
$\Lambda^*(1405)$ resonance emerges as a quasi-bound state of 
$N \overline{K}$ \cite{Kaiser}. 

\section{Summary and conclusions}

In this contribution we have analyzed the mass spectrum and the 
strong couplings of both strange and nonstrange baryons. The 
combination of a collective string-like $qqq$ model of baryons in 
which the orbitally excited baryons are interpreted as collective 
rotations and vibrations of the strings, and a simple elementary 
emission model for the strong decays can account for the main 
features of the data.  

The main discrepancies are found for the low-lying $S$-wave states,  
specifically $N(1535)$, $N(1650)$, $\Sigma(1750)$, $\Lambda^*(1405)$, 
$\Lambda(1670)$ and $\Lambda(1800)$. All of these resonances  
have masses which are close to the threshold of a meson-baryon 
decay channel, and hence they could mix with a quasi-molecular 
$S$ wave resonance of the form $qqq-q\overline{q}$. 
In contrary, decuplet baryons have no low-lying $S$  
states with masses close to the threshold of a particular decay 
channel, and their spectroscopy is described very well. 

The results of our analysis suggest that in future experiments 
particular attention be paid to the resonances mentioned above 
in order to elucidate their structure, and to look for evidence 
of the existence of exotic (non $qqq$) configurations of quarks 
and gluons.  

\section*{Acknowledgments}

This work was supported in part by DGAPA-UNAM under project IN101997  
and by CONACyT under project 32416-E.

\end{document}